\documentclass{article}
\usepackage{spconf,amsmath,graphicx}

\usepackage{enumitem}
\setlist{nosep, leftmargin=14pt}

\usepackage{hyperref}
\usepackage{fancyhdr}

\fancypagestyle{specialfooter}{%
  \fancyhf{}
  
  \fancyfoot[C]{\textbf{This work has been submitted to the IEEE for possible publication. Copyright may be transferred without notice, after which this version may no longer be accessible.}}
}

\title{Denoising Diffusion Probabilistic Models for Image Inpainting of Cell Distributions in the Human Brain}
\name{Jan-Oliver Kropp$^{1,2}$ \qquad Christian Schiffer$^{1,2}$ \qquad Katrin Amunts$^{1,3}$ \qquad Timo Dickscheid$^{1,2,4}$}
\address{
	$^{1}$ Institute of Neuroscience and Medicine (INM-1), Research Centre Jülich, Germany\\
	$^{2}$ Helmholtz AI, Research Centre Jülich, Germany\\
	$^{3}$ C\'{e}cile \& Oscar Vogt Institute for Brain Research, University Hospital Düsseldorf, Germany\\
	$^{4}$ Institute of Computer Science, Heinrich-Heine-University Düsseldorf, Germany
}

\usepackage[backend=biber,style=ieee]{biblatex}
\begin{document}
\thispagestyle{specialfooter}

\maketitle

\begin{abstract}
Recent advances in imaging and high-performance computing have made it possible to image the entire human brain at the cellular level.
This is the basis to study the multi-scale architecture of the brain regarding its subdivision into brain areas and nuclei, cortical layers, columns, and cell clusters down to single cell morphology
Methods for brain mapping and cell segmentation exploit such images to enable rapid and automated analysis of cytoarchitecture and cell distribution in complete series of histological sections.
However, the presence of inevitable processing artifacts in the image data caused by missing sections, tears in the tissue, or staining variations remains the primary reason for gaps in the resulting image data.
To this end we aim to provide a model that can fill in missing information in a reliable way, following the true cell distribution at different scales.
Inspired by the recent success in image generation, we propose a denoising diffusion probabilistic model (DDPM), trained on light-microscopic scans of cell-body stained sections.
We extend this model with the RePaint method to impute missing or replace corrupted image data.
We show that our trained DDPM is able to generate highly realistic image information for this purpose, generating plausible cell statistics and cytoarchitectonic patterns. 
We validate its outputs using two established downstream task models trained on the same data.
\end{abstract}


\section{Introduction}
\vspace*{-.5\baselineskip}
Advancements in imaging and high-performance computing have enabled whole section imaging of full tissue stacks of the entire human brain \cite{amunts2013bigbrain}, and allow for increasingly higher resolutions down to  micrometer resolution. This has stimulated the development of AI-driven methods for brain region segmentation~\cite{schiffer2021} and automated detection of cell bodies~\cite{upschulte2022contour}, which are key for linking such high-throughput imaging data to anatomical reference atlases~\cite{amunts2020julich}.
Nevertheless, the image data may contain artifacts caused by missing sections, tissue tears, or staining errors, resulting in gaps within the cell distributions.

Despite careful laboratory processes and quality control measures, such artifacts cannot be avoided in histological processing of large specimen. 
Based on the datasets considered in this work~\cite{amunts2013bigbrain}, we estimate the percentage of damaged or missing tissue between 5 and 7\% of the total tissue.
In order to provide complete cell distributions for analysis, imputation techniques are required to fill in the missing or corrupted data.
This imputation has, until now, been predominantly carried out manually or in a  semiautomated way at a resolution of $20 \mu{}m$~\cite{amunts2013bigbrain}. It has not been performed at a spatial resolution of $1 \mu{}m$, which is, however, prerequisite to 3D reconstruct histological data sets at cellular level. Our objective is to substitute manual labour with an automated method, utilising a deep learning model.

This model should optimally support downstream analyses such as cell detection~\cite{upschulte2022contour} and cytoarchitecture classification~\cite{schiffer2021}, by generating highly realistic image data which closely models the original cell distributions in terms of cell body sizes,  densities, and higher-level cytoarchitectonic features such as the laminar and columnar cortical organization. Furthermore, it should be capable for provenance tracking, to be reproducible.

\begin{figure}[t]
    \centering
    \includegraphics[width=\linewidth]{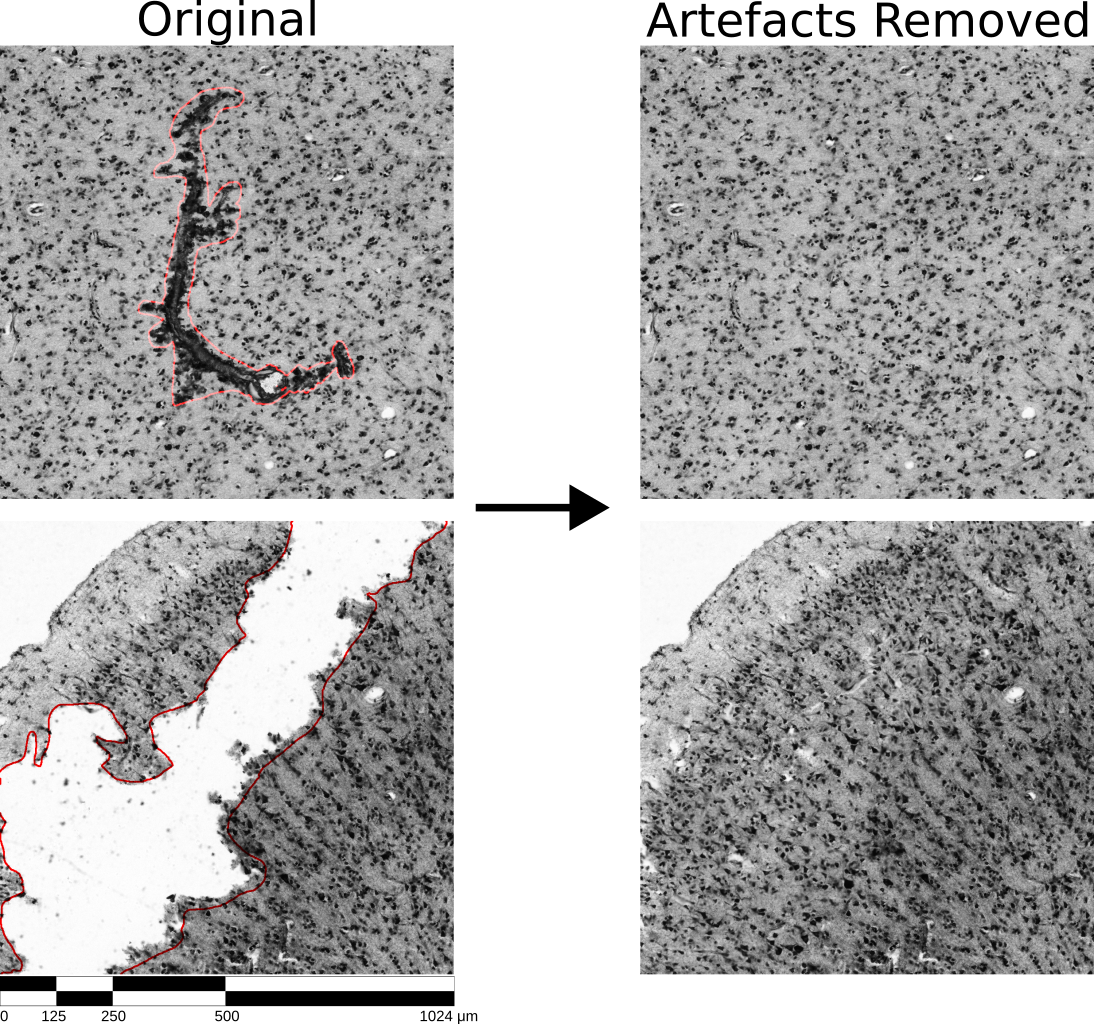}
    \caption{Example patches from the evaluation dataset with present artifacts. Left: Image patches with crystal artefact (top) and missing tissue (bottom). Right: Corresponding patches with artifacts removed by our DDPM and the RePaint method. Red annotations show manually provided annotations of the respective artefact. The repainted images show no signs of the artifacts and closely follow the true cytoarchitecture (e.g., the columnar organization).}
    \label{fig:introduction}
\end{figure}

To this end we propose an unconditional denoising diffusion probabilistic model~\cite{ho2020denoising} to replace previously labelled artifacts in image patches with realistic intact textures (Figure \ref{fig:introduction}).
We train the model on light-microscopic scans of cell-body stained histological human brain sections.
The unconditional model is extended with the RePaint algorithm~\cite{lugmayr2022repaint} to enable correction of artifacts without requiring re-training.
We evaluate the model's ability to mimic real data by analyzing the generated image data with two downstream neural networks trained for cell segmentation and cytoarchitectonic mapping, respectively. 

\section{Methods}
\vspace*{-.5\baselineskip}
\subsection{Denoising Diffusion Probabilistic Models}
Denoising diffusion probabilistic models~\cite{ho2020denoising} establish a relationship between noise and data distributions through an iterative denoising process, generating data samples from a noise source.
These models offer a probabilistic scheme for producing high quality data and have shown remarkable success in domains such as image synthesis and data imputation.
The implicit ability for self-correction in predictions during the stepwise diffusion process allows these models to learn intricate dependencies essential for modeling cell distributions, complexities challenging to capture with conventional approaches.
The model defines a forward and backward diffusion process.
During the forward process $q(x_t|x_{t-1})$, Gaussian noise is gradually added to the input image $x_0$, which here refers to a real image patch. It uses a predefined schedule which defines  number of iterations and noise level per iteration.
In the resulting output image $x_n$,  content is typically fully replaced by random noise. 
The output distribution then follows a normal distribution in image space.
The backward process $p_\theta(x_{t-1}|x_t)$ then reverts the stepwise distortions of the forward process. Other than the conceptually simple forward process, this is a complex inverse problem, which the DDPM approximates with a deep neural network. 
Following \cite{dhariwal2021diffusion}, we utilize a U-Net architecture for the backward process.
DDPMs are unconditional models, which means that they are designed to generate random samples of the image distribution. For our application however, we need to inform the model by the surrounding intact tissue so that it can produce textures capturing the specific properties of the local cell distributions.  
In this study, we condition the model using surrounding tissue information using the RePaint \cite{lugmayr2022repaint} approach.

\subsection{RePaint}
\vspace*{-.5\baselineskip}
The RePaint algorithm  expands on unconditional DDPMs to support inpainting of images, by altering the reverse diffusion process.
RePaint does not require any retraining or fine tuning and can readily be applied to the trained model.
In each step, the input to the DDPM is composed of the current unknown part of the image and the known part of the image with added noise of the current step
\begin{displaymath}\label{eq:repaint}
    x_{t-1} = p_{\theta}\left(M \cdot \tilde{x_t} + (1-M) \cdot x_t \right) \;,
\end{displaymath}
where $p_\theta$ is the aforementioned U-Net model and $M$ is the mask for the artefact or missing part. 
$\tilde{x_t}$ is the known image passed through the forward diffusion process at time step $t$ and $x_t$ is the currently generated image.
Through this approach, the newly generated part gets conditioned on the known part of the image.
Experiments in \cite{lugmayr2022repaint} showed the consistency of the generated part to the known part can be increased, by repeating each step $j$ times.
$j$ is also referred to as the jump length.
To repeat the step we again add noise to $x_{t-1}$ to create a new $x_t$ and use it in Equation \ref{eq:repaint}.
Due to the jumps, the model repeatedly combines known and unknown parts of the image for every diffusion step, increasing the dependency between the parts and generally improving the results.
We used a jump length of $j=5$ in the experiments. 

\subsection{Model architecture}
\vspace*{-.5\baselineskip}
As proposed by \cite{dhariwal2021diffusion}
we model $p_\theta$ as a deep residual U-Net with 14 residual blocks of layers, 7 for the encoding and 7 for the decoding part.
We added self-attention at resolutions 8, 16 and 32.
$q$ is modelled as a Gaussian forward diffusion process with cosine noise schedule for 256 diffusion steps.
The model was tasked to predict the mean and variance of the noise in each step.

\section{Results}
\vspace*{-.5\baselineskip}
\subsection{Dataset \& Training}
\begin{figure}
    \centering
    \includegraphics[width=\linewidth]{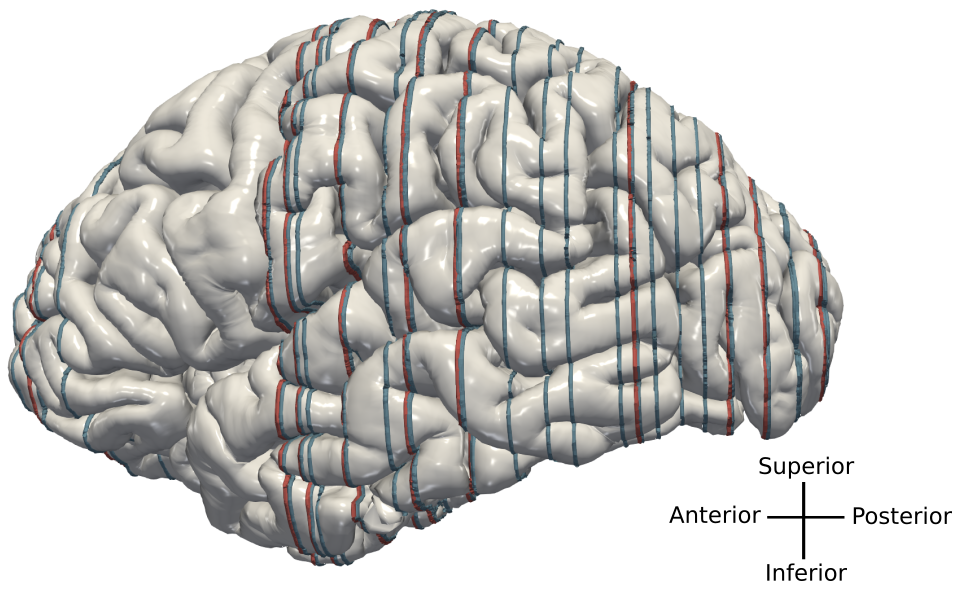}
    \caption{Approximate location of sections constituting to the training and evaluation datasets. Blue: Approximate locations of the training sections. Red: approximate locations of  evaluation sections.}
    \label{fig:datalocations}
\end{figure}

\vspace*{-.5\baselineskip}
Image patches for training and evaluation were sampled from 1 micron scans of the original tissue sections of the BigBrain model (\cite{amunts2013bigbrain}).
74 random brain sections were selected (similar to~\cite{sections1micron}, with 64 allocated for the training and 10 held back for evaluation.
Positions of training and evaluation sections are indicated in Figure \ref{fig:datalocations}.
For training, we sampled 20,000 patches randomly from gray matter regions in each of the 64 sections, resulting in 1.28 million samples.
The evaluation dataset was derived from the remaining sections and consisted of 10,000 patches with ~1000 samples per section.
The patch dimensions for all training and evaluation data were set to 1024x1024 pixels or $1 mm^2$ at a resolution of 1 micron, to facilitate the model's ability to learn larger cytoarchitectonic structures and patterns.
To exclude artifacts from all training and evaluation patches, we created an artifact mask for every section, masking all image parts which are depicting corrupted or missing tissue.
Training was conducted on 64 A100 GPUs on the supercomputer JURECA-DC~\cite{krause2018jureca} for 12 hours and 100,000 steps.
The learning rate was set to $1e-5$. All training and evaluation was performed using 16 bit floating precision.

\subsection{Unconditional Generation}
\begin{figure}
    \centering
    \includegraphics[width=\linewidth]{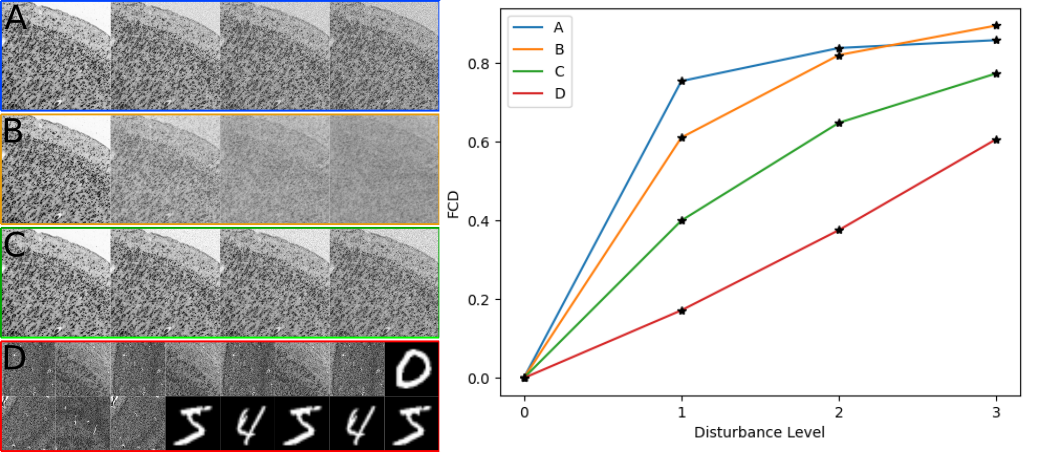}
    \caption{The FCD is evaluated for different disturbance levels of \textbf{A}: Gaussian Noise, \textbf{B}: Gaussian Blur, \textbf{C}: Salt \& Pepper Noise and \textbf{Dv}: Mixing the dataset with MNIST samples.}
    \label{fig:fcd-total}
\end{figure}

\vspace*{-.5\baselineskip}
Before addressing the repainting task, we analyzed the unconditional generative quality of the DDPM using the commonly used Frechét Inception Distance (FID)~\cite{heusel2017}. 
However, the FID is not well applicable to histology data, since we expect the cytoarchitectonic image data clearly outside the training data distribution of the Inception model. 
Instead, we utilize the latent space of a recently proposed cytoarchitecture classification model ~\cite{schiffer2021}.
The classification model is pretrained using a contrastive learning approach and then finetuned to classify brain areas based on their cytoarchitecture.
Just like the Inception model, it uses a latent space of size 2048.
To distinguish the newly proposed metric, we will refer to it as FCD (Frechét Cytoarchitecture Distance).
To demonstrate its efficacy in evaluating newly generated samples, we followed the procedure outlined in ~\cite{heusel2017} and assessed it against a range of perturbations.
The test results are presented in Figure \ref{fig:fcd-total}.
To derive the FCD value, we utilized 10,000 images from the evaluation set and 10,000 images generated by the DDPM.
The resulting FCD score for the model was $0.09$.
In contrast the FCD for an additional 10,000 images from the evaluation set generated a score of $0.006$.

\subsection{Artefact Repair}
\vspace*{-.5\baselineskip}
After evaluating the model in the unconditional setting, we investigated its practical performance in repairing tissue artifacts.
Artefact repair was run on a dataset of 10,000 image patches of size $1024 \times 1024$, sampled from the evaluation sections, which were not seen during training.
These patches were filtered for artifacts and can thus be used as intact reference data.
We obstructed random parts of the images using artefact masks from~\cite{liu2018image}. The area covered by the masks ranged from $\sim$ 5\% to 50\%.
The obstructed images were then repainted using the DDPM with the RePaint method.
Processing a patch of size $1 mm^2$ took 35 seconds on one A100 GPU.
We compared the image before and after repainting, measuring  cell statistics in the obstructed area and  cytoarchitectonic classification of the complete patch.
To compare the distortion introduced by the DDPM, we calculated the FCD of the repaired patches and the original intact patches, achieving a score of $0.0044$.

\subsubsection{Cell Statistics}
\begin{figure}
    \centering
    \includegraphics[width=\linewidth]{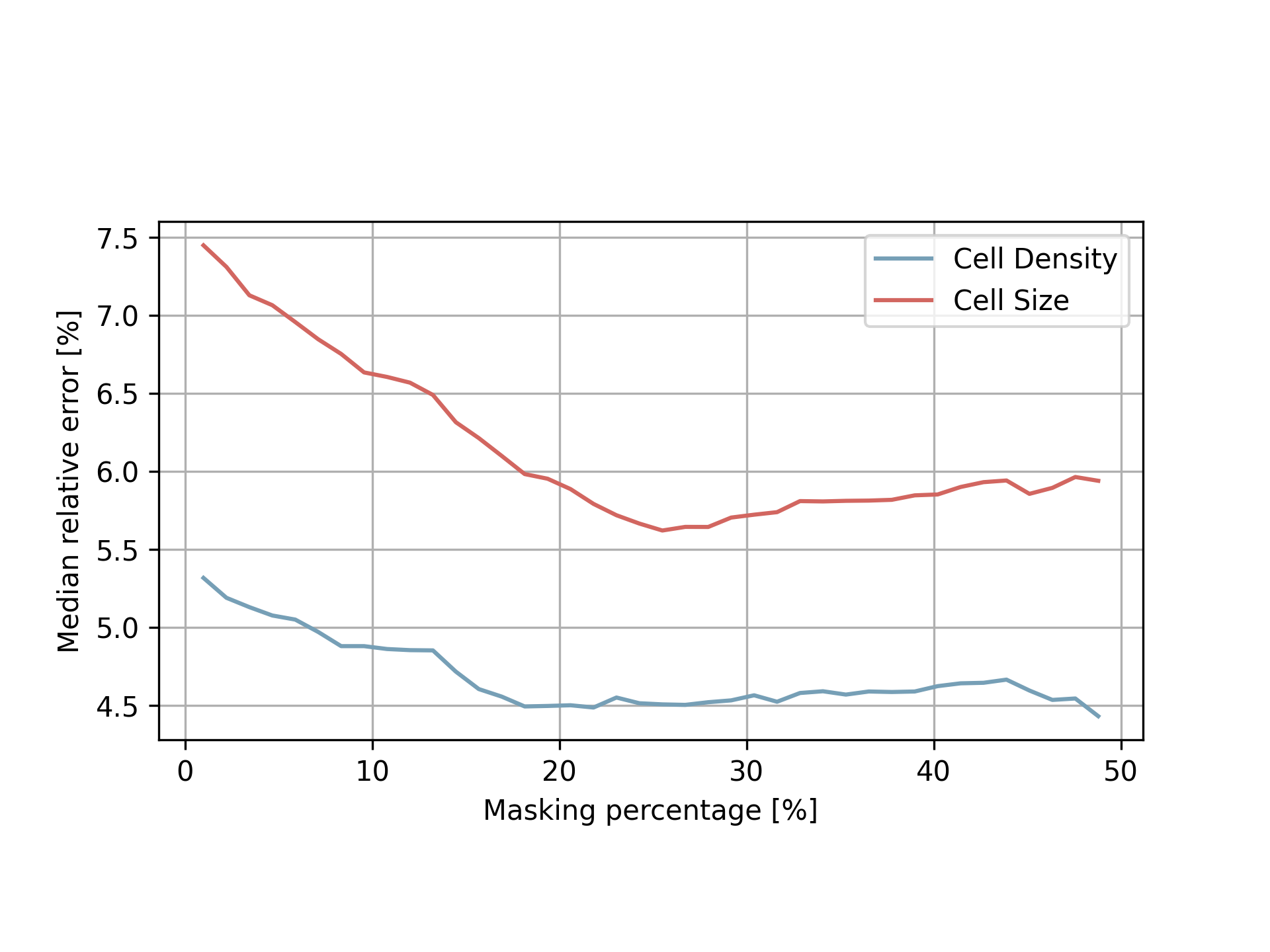}
    \caption{The median relative error in cell statistics after repainting with the proposed DDPM plotted against the masking percentage. Blue line: Median relative error in cell density. Red line: Median relative error in cell size. The median relative error for the cell size is below 7.5\% for all masking percentages, while the median relative error for density is below 5.5\%. }
    \label{fig:cellstatistics}
\end{figure}

\vspace*{-.5\baselineskip}
Brain sections stained for cell bodies can be used to evaluate cytoarchitectonic properties like densities, counts, sizes and shapes of cells.
We want to test whether the model preserves such fundamental attributes in an adequate fashion.
To this end we used the recently proposed Contour Proposal Network (CPN)~\cite{upschulte2022contour} for extracting the cell statistics in the masked region before and after the repainting.
As the CPN was trained on highly similar datasets, we  assume it to detect the the cell bodies sufficiently well, thus providing a suitable independent reference.   
The cell statistics arising from the images repaired by the DDPM were compared to the statistics extracted from the intact reference images. The median relative error was calculated for cell density (number of cells per area) and cell size (average size of the cells in a given area). The statistics were only compared in the masked regions.
The error was binned by percentage of tissue masked and the average and variance of these bins were matched against the percentage of masked tissue in figure \ref{fig:cellstatistics}.
The average relative error is below 10\% even for large obstructed regions.

\subsubsection{Cytoarchitectonic classification}
\begin{figure}
    \centering
    \includegraphics[width=\linewidth]{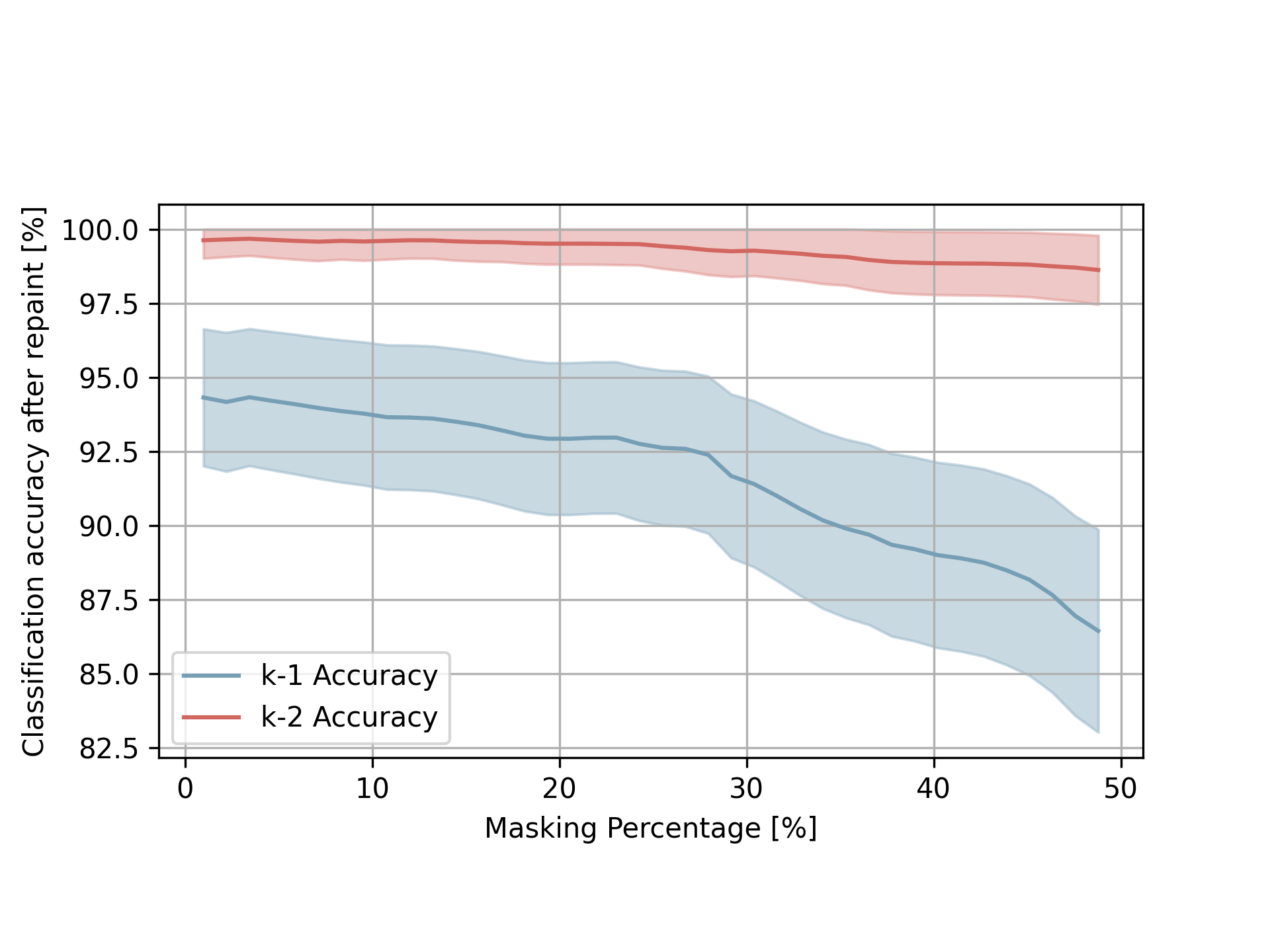}
    \caption{Relative number of patches where  predictions of an independent cytoarchitecture classification model \cite{schiffer2021} are consistent between intact and inpainted image patches according to $k=1$ and $k=2$ classification accuracy. Values are plotted over increasing size of the inpainted area. Lines represent the mean, underlying area represents the standard deviation.}
    \label{fig:classification-flips}
\end{figure}

\vspace*{-.5\baselineskip}
Cortical brain regions are characterized by their highly specific cytoarchitecture, for example relative thicknesses of the cortical layers.
These properties are essential for classifying brain regions, and should be preserved when inpainting missing information.  
To evaluate this effect, we use the cytoarchitectonic classification model proposed by~\cite{schiffer2021} to predict the brain region of intact and repaired image patches. 
If the model preserves essential cytoarchitectonic properties sufficiently well, the classification for a repaired patch should match the one for the original intact patch.
We tested classifications for 10,000 evaluation patches.
Figure \ref{fig:classification-flips} shows the percentage of patches where classification predicted the same brain region after repairing, averaged for different repainting area sizes.
To better capture possible inaccuracies in the classification model, we also compared the two most likely classifications after the artefact was repaired with the previous classification ("$k=2$ accuracy").
The graph shows a clear trend, where larger artefact constitute a higher percentage of wrong classifications after repainting. 
The overall average classification accuracy is above 85\%.
When we account for possible inaccuracies in the model classification, the $k=2$ accuracy is above 98\%.

\section{Discussion \& Conclusion}
\vspace*{-.5\baselineskip}
Visual inspection by experts and evaluation with the proposed FCD shows that the presented model is able to generate plausible high quality images for repainting artifacts, as well as for unconditional image generation.
Images generated by the inpainting model show consistent cell statistics, with deviations below 7.5\% regarding essential properties like cell density and size.
The model successfully repairs image artifacts by generating plausible structures, which are suitable for subsequent analysis algorithms, including cell detection and cytoarchitecture classification.
This way the model provides an important basis for enabling microstructural analysis of large tissue stacks with  missing or corrupted parts. 

To improve positioning and appearance of structures such as blood vessels, cortical columns or the pial surface, across brain sections, future work will consider how information from neighboring sections can be incorporated into the process, for example using Siamese neural networks~\cite{chopra2005}, and how to reduce the computational time required to repair artifacts.
For example, repairing artifacts in all sections of the BigBrain~\cite{amunts2013bigbrain} dataset would require approximately 645 days on a single A100 GPU, if the proportion of corrupt tissue is assumed to be $\approx 5\%$.
Stable diffusion~\cite{rombach2022high}, which applies diffusion in a smaller latent space, could help to reduce the computational costs significantly.
In addition, runtime could be further reduced by distilling the number of diffusion steps~\cite{meng2023distillation}.
Finally, employing methods for automatic artifact detection to replace manual annotation of artifacts (Figure \ref{fig:introduction}) will facilitate application of the proposed method at higher throughputs.

\section{Compliance with ethical standards}
\vspace*{-.5\baselineskip}
The brain used for this study was originally obtained in accordance to
legal and ethical regulations and guidelines as part of the body donor program of the Department of Anatomy of
the Heinrich Heine University Düsseldorf. The body donor (65 years old, male) gave written informed consent
for the general use of post-mortem tissue for aims of research and education. All usage in this work is covered by
a vote of the ethics committee of the Medical Faculty of the Heinrich Heine University Düsseldorf ($\#4863$).

\section{Acknowledgments}
\vspace*{-.5\baselineskip}
This work was funded by Helmholtz Association’s Initiative and Networking Fund through the Helmholtz International BigBrain Analytics and Learning Laboratory (HIBALL) under the Helmholtz International Lab grant agreement InterLabs-0015 and Priority Program 2041 (SPP 2041) “Computational Connectomics” of the German Research Foundation (DFG). The authors acknowledge the computing time granted through JARA on the supercomputer JURECA-DC at Forschungszentrum Jülich. We thank Tim Kaiser for valuable initial investigations as part of his master thesis.

\printbibliography

\end{document}